# Density functional theory study of rutile $VO_2$ surfaces

Thomas A. Mellan and Ricardo Grau-Crespo*

*Department of Chemistry, University College London, 20 Gordon Street, London WC1H 0AJ, U. Email: r.grau-crespo@ucl.ac.uk*

($24^{th}$ August, 2012)

**Abstract**

We present the results of a density functional theory (DFT) investigation of the surfaces of rutile-like vanadium dioxide, $VO_2(R)$. We calculate the surface energies of low Miller index planes, and find that the most stable surface orientation is the (110). The equilibrium morphology of a $VO_2(R)$ particle has an acicular shape, laterally confined by (110) planes and topped by (011) planes. The redox properties of the (110) surface are investigated by calculating the relative surface free energies of the non-stoichiometric compositions as a function of oxygen chemical potential. It is found that the $VO_2(110)$ surface is oxidized with respect to the stoichiometric composition, not only at ambient conditions but also at the more reducing conditions under which bulk $VO_2$ is stable in comparison with bulk $V_2O_5$. The adsorbed oxygen forms surface vanadyl species much more favorably than surface peroxo species.

## I. INTRODUCTION

At a temperature $T_c \approx 341$ K, a reversible first-order metal-semiconductor phase transition occurs in pure vanadium dioxide, $VO_2$.[1] The abrupt change in electrical resistivity is accompanied by a geometric distortion: in the state above $T_c$ the oxide has a tetragonal rutile-like structure ($VO_2(R)$), while it becomes monoclinic ($VO_2(M_1)$) below $T_c$. The transition from the semiconductor to the metallic state upon heating causes the transmittance in the near-infrared region to sharply decrease.[2] This makes vanadium dioxide of potential practical interest for 'smart' windows coatings, which can automatically regulate the temperature inside buildings by 'switching off' the transmission of the heat-carrying infrared radiation when the ambient temperature increases over a certain value.[3-5] The transition point of $VO_2$ can be adjusted to near room temperature via doping, for example, with 2 at.% tungsten.[6]

The mechanism of the transition has been debated for decades, and no clear answer has emerged yet, with some authors arguing that it is a Mott transition (dominated by correlation effects) and others that it is a Peierls transition (driven by the structural modification) or a combination of the two mechanisms.[7-10] Modelling this material based on first principles is challenging, and it is not clear whether any form of band theory can describe correctly the electronic structure of the system, especially the low-temperature phase. Both the local density approximation (LDA) and the generalized gradient approximation (GGA) of the density functional theory (DFT) fail to reproduce the finite band gap of the $VO_2$(M1) structure.[7,9] Eyert[11] has recently shown that the band gap in the M1 phase can be reproduced if the exchange term is calculated using a mixture of GGA and Hartree-Fock contributions, as implemented in the Heyd-Scuseria-Ernzerhof (HSE) functional.[12] However, more recent calculations have shown that the HSE description of $VO_2$ is also problematic: it predicts a wrong magnetic groundstate for the M1 phase, and also an artificial band gap opening in the groundstate of the R phase.[13]

This article presents a theoretical investigation of the stability and local geometry of low-index $VO_2$(R) surfaces, and the resulting equilibrium morphology of $VO_2$(R) particles. Understanding the properties of $VO_2$ surfaces is important in the context of thermochromic applications. Experimental research has shown that the morphology and preferred crystallographic orientation of $VO_2$ particles in thin films can have significant effects on both the hysteresis width and the critical temperature of the transition.[14] Furthermore, the study of the free surfaces is a necessary first step for the future investigation of the interactions of $VO_2$ with metal nanoparticles, which are able to modify the optical properties of the thermochromic films,[15,16] and of the interfaces of $VO_2$ with the glass substrate and other coating layers, like antireflective $TiO_2$. The rutile phase $VO_2$(R) is chosen for our study because it is the one stable under typical growth/synthesis conditions (e.g. chemical vapor deposition of $VO_2$ thin films is generally done at done at temperatures above 800 K).[17,18] Since this phase does not exhibit a band gap, the local DFT approach is not as problematic here as in the description of the $VO_2(M1)$ phase. We also discuss the redox behavior of the most stable surface of $VO_2(R)$ as a function of temperature and oxygen partial pressure.

## II. METHODOLOGY

Periodic DFT calculations were performed with the VASP program,[19,20] using the GGA approximation in the form of the Perdew-Burke-Ernzerhof (PBE) exchange-correlation functional.[21] The projected augmented wave method[22,23] was used to describe the interaction between the valence electrons and the core, and the core states (up to 3p in V and 1s in O) were kept frozen at the atomic reference states. The number of planewaves was determined by a cutoff kinetic energy of 500 eV, and the mesh of k-points for the bulk unit cell calculations contained 6x6x9 divisions. These parameters were checked with respect to convergence of the bulk total energy. For surface calculations, the number of k-points was adapted to achieve a similar sampling density in the reciprocal space.

As in previous DFT investigations of $VO_2(R)$,[7,9,11] our calculations are not spin polarized. However, contrasting with $VO_2(M1)$, where $V^{4+}$-$V^{4+}$ dimerization leads to spin pairing,[24] in $VO_2(R)$ $V^{4+}$ cations are expected to have a single unpaired $3d$ electron, and there is experimental evidence of paramagnetic behavior of the high-temperature phase.[25] Although a non-magnetic calculation is not a good representation of a paramagnetic phase, a spin-polarized calculation with ordered magnetic moments is not generally adequate either. We have tried nonmagnetic (NM), ferromagnetic (FM) and antiferromagnetic (AFM) calculations of the $VO_2(R)$ bulk using the GGA-PBE functional, and found that the FM groundstate exhibits a gap at the Fermi level in the minority spin channel (figure 1), and metallic behavior of the majority spin electrons. This strong spin asymmetry means that the FM groundstate is a poor representation of the paramagnetic state of $VO_2(R)$. On the other hand, while an AFM solution can lead in principle to a more realistic symmetric behavior of majority and minority spin electrons, our GGA calculation with initial AFM ordering of the moments converged to a solution very close to the NM groundstate, with nearly zero final magnetic moments. Spin-polarized solutions with different orientation of the magnetic moments can be stabilized for $VO_2(R)$ using the screened hybrid functional HSE, but as mentioned above, this leads to solutions with artificial gaps at the Fermi level (in both the majority and minority spin channels),[13] which is in conflict with the well-established metallic character of $VO_2(R)$.[26,27] Therefore, in the present work we use the plain GGA functional, without any Hartree-Fock exchange contribution, which gives the correct metallic character of this phase.

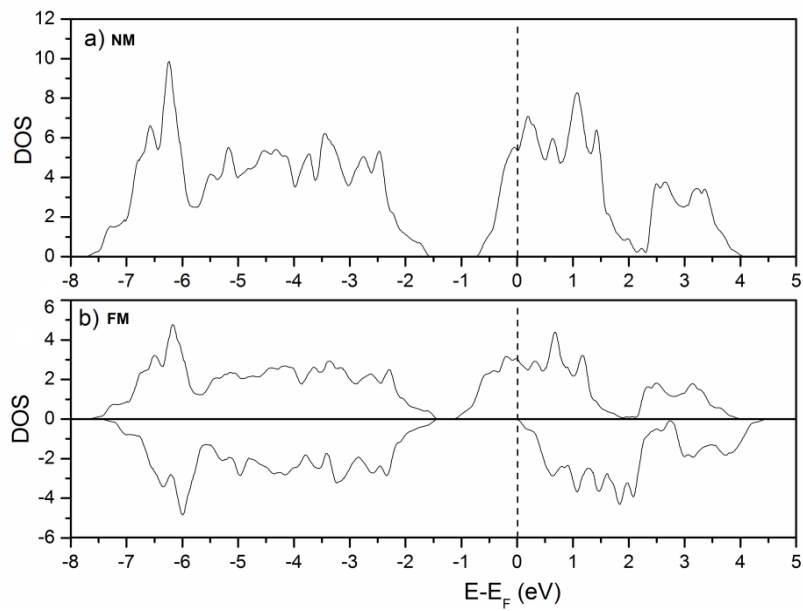

**FIG 1. Electronic density of states (DOS) for the $VO_2(R)$ bulk a) without spin polarization, and b) with ferromagnetic order of the magnetic moments.**

We first consider the stabilities of five low-index surface orientations ((110), (010), (001), (111) and (011)) by performing periodic calculations in slabs with stoichiometric composition, thicknesses between 11 and 14 Å (depending on the orientation), and vacuum gaps of ~14 Å (figure 2). The two surfaces of each slab are symmetrically equivalent, and this equivalence is kept during all of the calculations, preventing the formation of the electric dipole moments that can be associated with asymmetric slabs. The cell parameters of the slab are kept constant during the calculations, based on the relaxed cell parameters of the bulk: $a$=4.617 Å and $c$=2.774 Å, which are in reasonable agreement with experimental values $a_{exp}$=4.554 Å and $c_{exp}$=2.857 Å (deviations of +1.4% and -2.9% for $a$ and $c$, respectively, and of -0.1% in the cell volume).[28] All atoms in the slab are fully relaxed, and the surfaces energies are then obtained using the standard expression:

$$\gamma = \frac{E_{slab} - E_{bulk}}{2A}, \qquad (1)$$

where $E_{slab}$ is the energy per slab unit cell, $E_{bulk}$ is the energy of an equivalent amount of bulk solid, and $A$ is the surface area. The equilibrium morphology of a $VO_2$ (R) particle (ignoring higher Miller indices) is constructed using Wulff's method, which requires that the distance to a given surface from the center of the particle is proportional to the surface energy.[29]

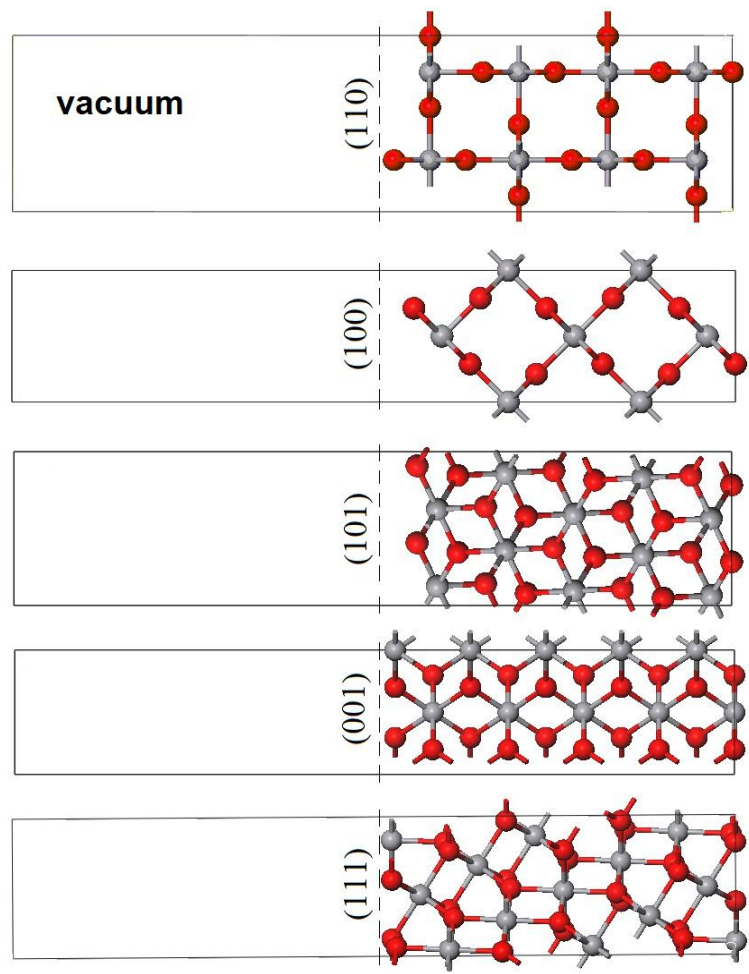

**FIG.2. Slab models for the low-index surfaces of $VO_2$ (R). The surfaces are shown before relaxation.**

We examine the most stable surface further by removing or adding oxygen atoms to form non-stoichiometric compositions. Care is taken here to preserve the symmetrical equivalence of the two surfaces of the slab when removing or adding atoms. The discussion of the stabilities of non-stoichiometric surface terminations is based on the ab initio thermodynamics formalism first introduced by Scheffler et al..[30] The surface free energy is calculated as:

$$\sigma(T,p) = \frac{E_{slab} - E_{bulk}}{2A} - \frac{\Gamma}{A}\mu_O(T,p) \tag{2}$$

where

$$\Gamma = \frac{1}{2}(N_O - 2N_V) \tag{3}$$

is the excess number of O ions at each surface of the slab ($N_O$ and $N_V$ are the numbers of O and V ions in the slab model, respectively). It is possible to express the chemical potential of oxygen, assuming equilibrium with the gas phase, as:

$$\mu_O(T,p) = \frac{1}{2}\left(E[O_2] + \Delta g_{O_2}(T,p_0) + k_B T \ln\frac{p}{p_0}\right) \tag{4}$$

The first term within the bracket is the DFT energy of the oxygen molecule. The second term is the difference in the Gibbs free energy per $O_2$ molecule between 0 K and $T$, at $p_0$=1 bar; this contribution can be extracted from thermodynamic tables[31] in order to avoid the explicit simulation of the gas phase, as done in previous studies.[32,33] The last term represents the change in free energy of the oxygen gas when the pressure changes from $p_0$ to $p$ at constant temperature $T$, assuming ideal gas behavior. We follow the usual convention of expressing the oxygen chemical potential with reference to half the energy of the $O_2$ molecule, that is:

$$\mu_O(T,p) - \frac{1}{2}E[O_2] \to \mu_O(T,p), \tag{5}$$

which makes the chemical potential independent of calculated quantities. In the evaluation of the slab energies we then must subtract half of the energy of the $O_2$ molecule for each oxygen atom in the slab, for consistency. With this method, it is possible to plot the surface free energies given by equation (2) for different surface compositions as a function of chemical potential, and discuss the redox behavior of the surface. More details about the method can be found elsewhere.[32,33]

## III. RESULTS AND DISCUSSION

### A. Stoichiometric Surfaces

The surface energies for the five crystallographic planes shown in Figure 2 are listed in Table I. We are not aware of previous calculations or experimental determinations of surface energies and geometries of rutile $VO_2$, so we compare our results with those reported for isostructural rutile $TiO_2$, whose surfaces are very well characterized.[34] As for rutile $TiO_2$,[35] the most stable surface of rutile $VO_2$ is the (110), with a calculated surface energy of only 0.29 $J/m^2$.

Table I. Calculated surface energies for low-index surfaces of $VO_2(R)$.

| Surface | $\gamma$ ($J/m^2$) |
| --- | --- |
| (110) | 0.29 |
| (100) | 0.42 |
| (101) | 0.75 |
| (001) | 0.96 |
| (111) | 1.25 |

The $VO_2$(110) surface terminates in bridging oxygen atoms forming rows along the [001] direction, followed by a plane which includes triple-coordinated oxygen atoms, and fivefold- and sixfold-coordinated vanadium atoms. It relaxes very similarly to the $TiO_2$(110) surface.[34] The main relaxations occur perpendicularly to the surface, with the fivefold-coordinated vanadium atoms relaxing inward and the sixfold-coordinated vanadium atoms and the in-plane oxygen atoms relaxing outward, as shown in figure 3. The surface plane of fivefold V atoms ends up being ~0.42 Å below the plane of sixfold V atoms. This is again similar to the relaxed $TiO_2$(110) surface, where the equivalent distance is 0.44 Å, according to low-energy electron diffraction experiments (and 0.39-0.42 Å in theoretical calculations, depending on the method).[36]

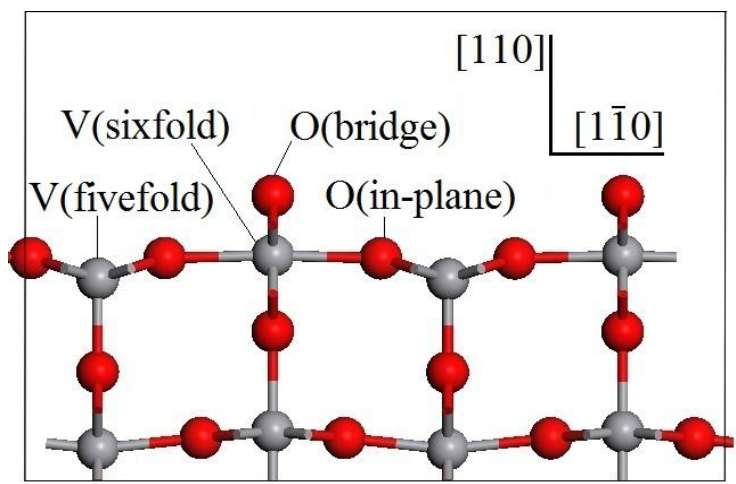

**FIG.3. The relaxed VO$_2$(R)(110) with stoichiometric composition, showing the notation for the surface atoms.**

The order of stabilities of VO$_2$(R) surfaces is (110)>(100)>(101)>(001)>(111) (Table I), and it also agrees with what has been reported for rutile TiO$_2$ (e.g. in reference [37], although the TiO$_2$(111) surface was not considered there). When the surface energies in Table I are used to construct the equilibrium morphology of VO$_2$(R), the acicular shape shown in figure 4a is obtained. As expected, the (110) plane dominates the morphology. The (100) does not appear in the morphology, despite being the second most stable surface, because the ratio $\frac{\gamma(100)}{\gamma(110)} = 1.45 > \sqrt{2}$ implies that the (100) planes are cut out by the very stable (110) planes in the Wulff's construction. In order to be present in the morphology, the surface energy of the (100) surface would need to be $\gamma(100) < \sqrt{2}\gamma(110)$ (figure 4b). The other surface appearing in the equilibrium morphology of VO$_2$(R) is the (101), at the top and bottom of the acicular shape.

Experimental crystal habits of VO$_2$ have been described by several authors. Misra et al.[24] reported two types of crystal habits, both exhibiting the (110) surface prominently, but one more elongated along the [001] direction than the other. The cross section of the particles was found to be a square limited by (110) surfaces, as in our calculations, although the termination of the particle at the top and bottom of the acicular shape is not the same as in our theoretical equilibrium morphologies (the termination planes are not identified in this experimental study). Sohn et al.[38] and Löffler et al. have described $VO_2$ nanowires which grow along the rutile [001] direction, and exhibit the (110) surface prominently. These nanostructures are grown at high temperatures but characterized at low temperatures, when their crystal structure becomes monoclinic. It is reasonable to expect that the experimental crystal habit is not modified by the metal to semiconductor transition, which structurally consists of the dimerization of V-V cations along the rutile $c$ axis; therefore it is valid to compare our equilibrium morphology for $VO_2(R)$ with the morphologies observed at low temperature in monoclinic samples. It should be noted that equivalent crystal directions have different indexes in the rutile and the monoclinic structures: the rutile [001] direction corresponds to the monoclinic [100] direction, while the rutile (110) surface corresponds to the monoclinic (011) surface.

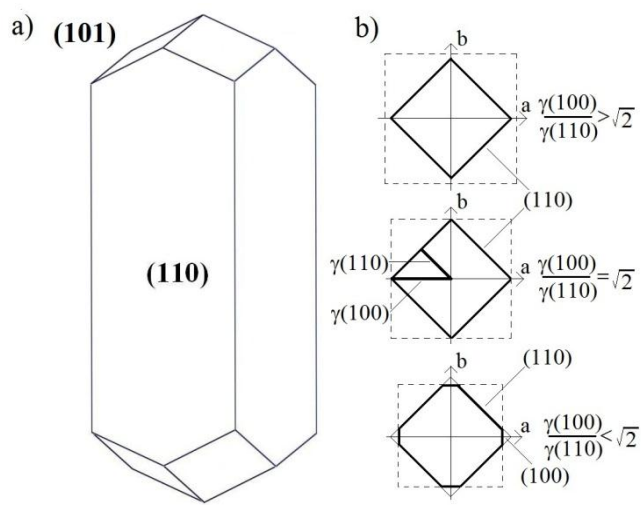

**FIG. 4. a) Wulff's construction of the equilibrium morphology for a $VO_2$ (R) particle; b) scheme of the cross-section of the particle for different ratios of stabilities of the lateral surfaces, illustrating why the (100) surface is absent in the equilibrium morphology.**

## B. Redox behavior of the (110) surface

We now discuss the redox properties of the most stable surface of $VO_2(R)$, the (110), by comparing the surface free energies corresponding to different oxygen to vanadium ratios at the surface. The number of vanadium atoms in the slab is kept the same as in the stoichiometric surface, but the number of oxygen atoms at each surface is changed by Γ (given by equation 3). Because of the size of our supercell, and assuming that oxygen atoms occupy bulk-like positions around the surface vanadium atoms, only five values of Γ are possible if we constraint to a maximum of 1 monolayer (ML) of adatoms or vacancies: Γ=0 surface is the stoichiometric surface, Γ=1, 2 are the partially and totally oxidized surfaces, and Γ=-1, -2 are the partially and totally reduced surfaces (figure 5). Oxygen positions other than the bulk-like sites will be considered in the case of the fully oxidized surface. Of course, intermediate degrees of oxidation and reduction could be also investigated by using a larger supercell and a configurational analysis of the distribution of excess oxygen/vacancies in order to get a more quantitative picture of the redox behavior, but we will see that the present "coarse-grained" picture already provides very useful information.

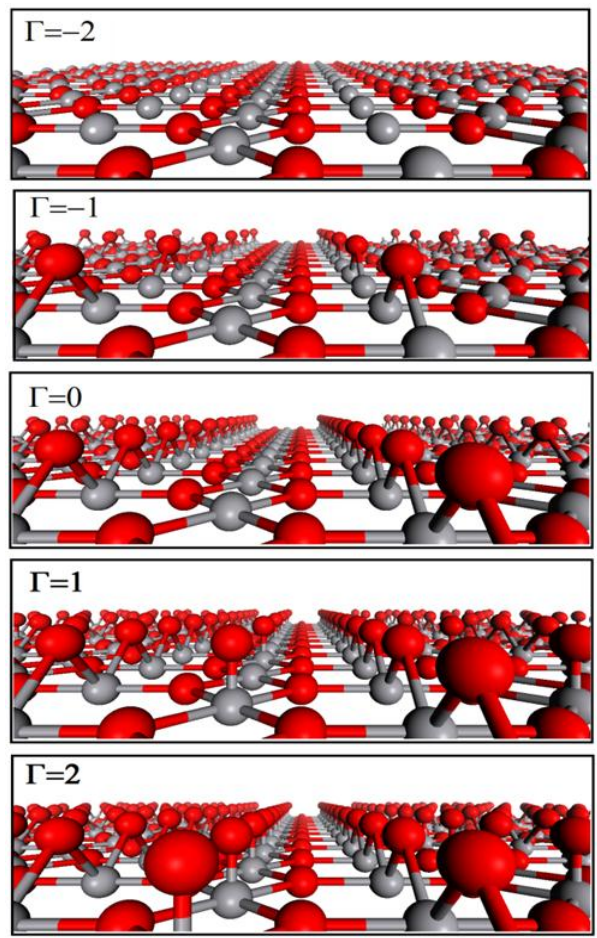

**FIG. 5. The $VO_2$ (R)(110) with different amounts of surface oxygen. Γ=0 surface is the stoichiometric surface; Γ=-1, -2 are the partially and totally reduced surfaces; Γ=1, 2 are the partially and totally oxidized surfaces, all shown before relaxation. Oxygen atoms are red and vanadium atoms are grey.**

For the analysis that follows we need the energy of an $O_2$ molecule. DFT calculations of the triplet groundstate yields a binding energy of -6.08 eV (with respect to triplet oxygen atoms) and an equilibrium bond length of 1.23 Å, which compares well with previous computational studies.[39,40] The experimental value for the binding energy at absolute zero (-5.12 eV)[41] is considerably less negative than the theoretical result. The overbinding of the $O_2$ molecule by GGA calculations (by 0.96 eV in our calculations) is a well-known effect.[21] Wang et al.[39] have argued that, besides the $O_2$ overbinding (and the correlation errors in the description of the *d* orbitals in the case of transition metal cations), there is an additional error in the calculation of the energy of redox reactions involving oxides, which is associated with adding electrons to the O 2p orbitals upon the formation of lattice $O^{2-}$ species. These authors therefore proposed a correction based on the average shift required in the energy of the oxygen molecule in order to reproduce the formation energy of several oxides from their metals. We follow a similar procedure here, which is illustrated in figure 6, using the formation energies of five tetravalent oxides: $HfO_2$, $ZrO_2$, $TiO_2$, $SiO_2$ and $GeO_2$. An upward shift of 1.05 eV in the $O_2$ energy (making it less negative) brings the experimental and theoretical formation energies into good agreement for all oxides. Contrasting with the result in ref. [39], in our case this shift is very similar to the magnitude of the $O_2$ overbinding in the GGA, which implies that most of the systematic error in the oxide formation enthalpies comes from the $O_2$ overbinding, while the error associated with the filling of the O 2p orbitals in the oxides is very small. The difference between the present results and those in [39] probably arises from us only considering tetravalent oxides in our analysis, while they mainly used alkaline and alkaline earth metal oxides. In what follows, redox energies will be reported with respect to both the uncorrected and the corrected energy of the oxygen molecule.

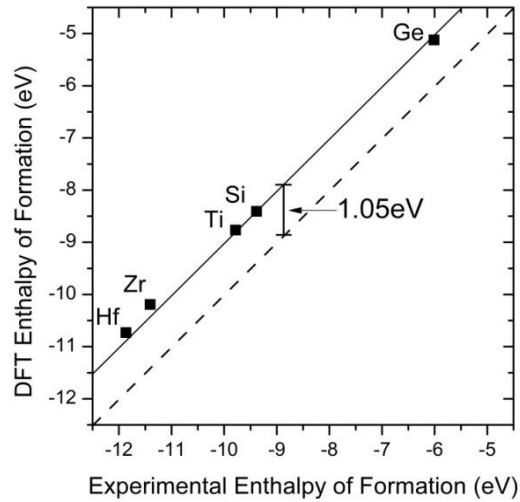

**FIG. 6. Experimental enthalpies of formation of metal oxides plotted against the GGA values, showing the necessary correction due to O$_2$ over-binding in the GGA.**

We first discuss the reduced surfaces. Removing one oxygen atom at each surface of the slab, leads to the surface with Γ=-1 (0.5 ML of oxygen vacancies). The vacancy formation energy, calculated as:

$$E_{\text{vac}} = {}^1\!/_2 \left( E[O_2] + E[\Gamma = -1] - E[\Gamma = 0] \right) \tag{6}$$

is 3.33 eV (3.85 eV after correction) for the bridging oxygen vacancies and 4.30 eV (4.83 eV after correction) for the in-plane oxygen vacancies. The higher stability of the bridging vacancy with respect to the in-plane vacancy has also been reported for the TiO$_2$(110) surface.[42-44] The vacancy formation in the VO$_2$ bulk is 3.00 eV (3.53 eV after correction), which suggest that any surface vacancies will tend to migrate towards the bulk. This contrasts with TiO$_2$ rutile, where vacancies are significantly easier to create in the (110) surface (in bridging positions) than in the bulk.[44] Removing the rest of the bridging oxygen atoms from the partially reduced VO$_2$(110) surface, forming the surface with Γ=-2, takes much more energy per vacancy (4.07 eV, or 4.60 eV after correction) than removing the first bridging oxygen atoms, but still less than removing in-plane oxygen atoms. This indicates that in-plane oxygen vacancies are only created after all the bridging oxygen atoms have been removed from the surface. We performed a Bader analysis[45] of the variation in the charges of the surface ions upon the creation of an oxygen vacancy in a bridging position. The positive charge of all the surface cations is slightly reduced, but most of the reduction is localized on the two V ions that were bridged by the removed oxygen, for which the Bader charge went from +1.88$e$ in the stoichiometric surface to +1.71$e$ in the partially reduced surface ($e$ is the elementary positive charge). The charges of the other surface V cations decreased by less than 0.05$e$ each. This can be loosely interpreted as the formation of $V^{3+}$ cations upon the formation of oxygen vacancies, but it should be remembered that this is metallic system with strong charge delocalization, and therefore any interpretation of the redox behavior in terms of ionic formal charges is very approximate.

On the other hand, the adsorption of one oxygen atom at the surface, leading to 0.5 ML of adatoms ($\Gamma$=+1), involves the energy

$$E_{\text{ads}} = {}^{1}\!/_{2}\,(E[\Gamma = +1] - E[\Gamma = 0] - E[O_2]) \qquad (7)$$

per adatom, which is -1.58 eV (-2.10 eV after correction). We are assuming here that each oxygen adatom goes on top of a fivefold coordinated V atom, forming a vanadyl species; other configurations will be discussed below. The fact that oxygen adsorption from the gas phase is strongly exothermic already suggests that surface oxidation will be thermodynamically favorable, but a complete analysis requires consideration of the gas partial pressure in equilibrium with the surface. Completing 1 ML of vanadyl species (forming a surface with $\Gamma$=+2, from the surface with $\Gamma$=+1) is also an exothermic process, involving -1.03 eV (-1.56 eV after correction) per adatom. The vanadyl V=O equilibrium bond length is 1.61 Å, which compares well with values reported elsewhere for surface vanadyl species (1.55-1.70 Å).[46] Due to the formation of a double bond, this bond length is much shorter than the V-O distance in the bulk (1.95 Å in our calculations). The Bader analysis indicates that the vanadyl oxygen charge (-0.51$e$) is less negative than the bridging oxygen, whose charge goes from -0.76$e$ in the stoichiometric surface to -0.71$e$ in the oxidized surface. Upon oxidation, the charges of all the surface cations increase by a similar small amount (~0.05$e$).

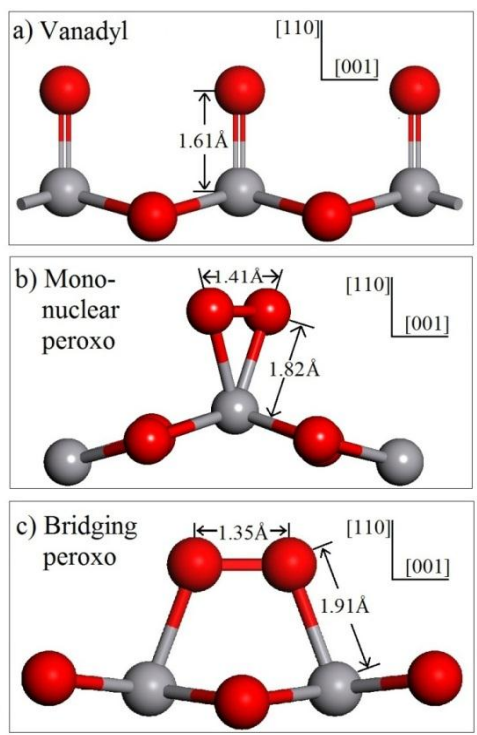

**FIG. 7. Different configurations with oxygen excess Γ=2: a) vanadyl terminated, b) peroxo group coordinated to one V atom, c) peroxo group bridging between two V atoms. Relaxed geometries are shown. O atoms are red and V atoms grey.**

The vanadyl-terminated surface is not the only surface that can be formed with Γ=+2. Oxygen can be adsorbed also as peroxo species $(O_2)^{2-}$, with less electron transfer from the surface vanadium atoms to the adatoms than in the case of vanadyl formation. The presence of peroxo species has been investigated for other oxide surfaces, e.g. in the surfaces of alkaline-earth oxides,[47] in $MoO_3$(010),[48] in $V_2O_3$(0001),[49] and in $FeSbO_4$(100).[32] We have investigated here two types of surface peroxo configurations, which are shown in figure 7: a flat peroxo species coordinated to one V atom (figure 7b), and a peroxo species bridging two V atoms (figure 7c). We find that both peroxo configurations are locally stable, that is, they correspond to potential energy minima. The O-O distances of 1.41 Å and 1.35 Å compare well with theoretical values reported elsewhere for surface peroxo groups, [32,47-49] and are slightly below the experimental O – O distance of the peroxo ion in hydrogen peroxide (1.475 Å).[41] The Bader analysis indicates that the charge of the oxygen atoms forming the peroxo groups (~-0.3$e$) is significantly less negative than the lattice oxygen species at the surface, as expected. The mono-coordinated peroxo species is slightly more stable than the bridging species (by ~24 meV per peroxo), but it is much higher in energy than the vanadyl-terminated surface discussed earlier (by 1.72 eV per peroxo). Therefore, must of the excess oxygen at the $VO_2$(110) surface is expected to form vanadyl species.

Finally, we discuss the thermodynamics of surface reduction/oxidation as a function of temperature and oxygen partial pressure in the gas phase. Figure 8 shows the variation of the surface free energies for different compositions with the chemical potential (only the most stable configuration for each $\Gamma$ is used here). The chemical potential is plotted in terms of temperature and partial pressure of $O_2$ in the graph below, and given along abscissas for easy comparison with the top plot. To put these chemical potentials in context, we draw a vertical line at the value below which bulk $VO_2$ becomes stable with respect to bulk $V_2O_5$. This is approximately equal to the enthalpy of the reaction:

$$2VO_2 + \tfrac{1}{2}O_2 \rightarrow V_2O_5, \qquad (8)$$

$\Delta H$= -1.36 eV, which we obtain from the experimental formation energies of the two oxides.[50] At the chemical potential $\mu_O=\Delta H$, the free energy of the above reaction changes sign (we are neglecting here the small contributions from the variation of enthalpies with temperature and from the difference in entropy between the two oxides). The area to the left of the vertical line in the chemical potential plot thus corresponds to the conditions under which bulk $VO_2$ is thermodynamically stable with respect to bulk $V_2O_5$. At ambient conditions $V_2O_5$ is the stable bulk phase, while the synthesis of $VO_2$ requires high temperatures or a controlled oxygen atmosphere.

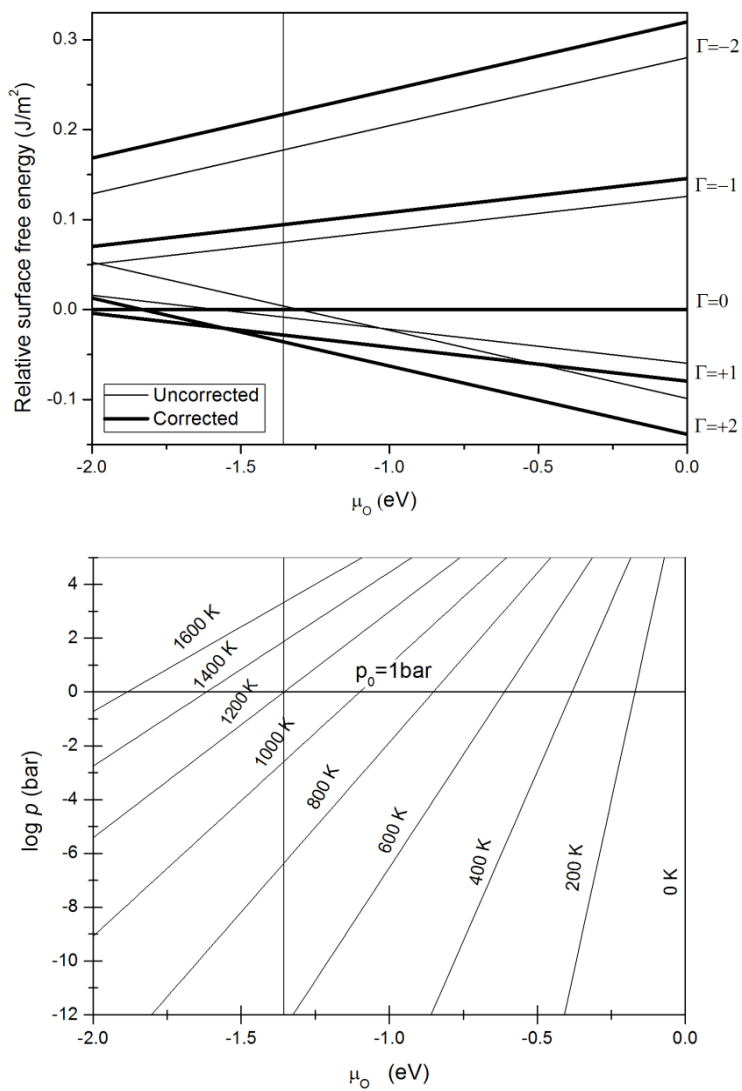

**FIG. 8. Top: relative surface free energies for different compositions of the VO$_2$(110) surface as a function of oxygen chemical potential. Bottom: chemical potential of oxygen in the gas phase as a function of temperature and oxygen partial pressure. The area to the left of the central vertical line represents the conditions under which bulk VO$_2$ is thermodynamically stable with respect to V$_2$O$_5$.**

The correction to the O$_2$ energy does not change the gradient of the lines plotted in figure 8(top), but does affect the relative positions of the lines (whereas the lines in the bottom of figure 8 are determined only by experimental information and are independent from the calculations). Regardless of whether the correction is applied or not, the fully oxidized surface is the most stable at ambient conditions. Surface oxidation of VO$_2$ can be expected for all the conditions under which V$_2$O$_5$ is the most stable bulk phase, including most temperatures and oxygen partial pressure of practical interest for applications. On the other hand, the predicted behavior at the very reducing conditions (below $\mu_O \approx -1.36$ eV) under which bulk $VO_2$ is stable with respect to $V_2O_5$ depends somewhat on whether the oxygen correction is introduced, as the uncorrected DFT calculations underestimate the level of oxidation. The corrected results suggest that there is surface oxidation even at these very reducing conditions. At ambient pressure, the excess surface oxygen forming vanadyl species becomes thermodynamically unstable with respect to a stoichiometric surface only at temperatures above ~1600 K.

The strong trend towards oxidation of the $VO_2$ surface seems consistent with the presence of $V^{5+}$ ions at the surface of $VO_2$ films, which was suggested on the basis of x-ray photoelectron spectroscopy (XPS) measurements by Manning et al..[6] However, according to these authors, a very thin film (10-20 nm) of $V_2O_5$ develops at the surface of $VO_2$ samples which are exposed to atmospheric conditions, and it is probably this surface film which leads to the $V^{5+}$ XPS signal. This experimental observation means that surface oxidation of $VO_2$ samples occurs at a level much deeper than what we have considered in our models. Our calculations therefore only describe the first stage of surface oxidation, after the formation of $VO_2$ crystals. Future work will be thus devoted to understand the development of $VO_2/V_2O_5$ interfaces.

## IV. CONCLUSIONS

As in rutile $TiO_2$, the oxygen-terminated (110) surface is the most stable in rutile $VO_2$. The surface relaxation patterns are also very similar in both oxides, as it is the order of stability of other low-index surfaces. The equilibrium morphology of $VO_2(R)$ has been found to be acicular, laterally confined by the (110) planes; therefore the formation of $VO_2$ nanowires growing along the rutile [001] direction is energetically favorable. Because the (110) surface is by far the most prominent surface in the $VO_2$ morphology, which is in agreement with experimental observations, future theoretical and experimental studies of the $VO_2$ surface behavior will probably focus on this surface.

The $VO_2(R)$ (110) surface can be expected to be oxidized even at strongly reducing conditions; it would then be very difficult to grow $VO_2$ samples that maintain the bulk stoichiometry at the surface. In the initial state of oxidation, the excess oxygen forms surface vanadyl groups, while peroxo species are comparatively much less stable. On the other hand, reduction of the surfaces is thermodynamically very unfavorable, and oxygen vacancies are in fact easier to form in the bulk than in the surface.

The application of a correction to the overbinding of the oxygen molecule within the GGA approximation is found to have a non-negligible effect on the prediction of the surface composition under reducing conditions. The uncorrected results indicate that the $VO_2$ surface will be stable against oxidation for oxygen chemical potentials in the region where bulk $VO_2$ is stable. However, after applying a correction to the oxygen molecule energy, it is found that much of the surface excess oxygen can actually survive these very reducing conditions.

**ACKNOWLEDGMENTS**

This research was funded by EPSRC grant EP/J001775/1. Via our membership of the UK's HPC Materials Chemistry Consortium, which is funded by EPSRC (EP/F067496), this work made use of the facilities of HECToR, the UK's national high-performance computing service, which is provided by UoE HPCx Ltd at the University of Edinburgh, Cray Inc. and NAG Ltd, and funded by the Office of Science and Technology through EPSRC's High End Computing Programme.